\PassOptionsToPackage{unicode}{hyperref}
\PassOptionsToPackage{hyphens}{url}
\documentclass[
  12pt,
]{article}
\usepackage{amsmath,amssymb}
\usepackage{iftex}
\ifPDFTeX
  \usepackage[T1]{fontenc}
  \usepackage[utf8]{inputenc}
  \usepackage{textcomp} 
\else 
  \usepackage{unicode-math} 
  \defaultfontfeatures{Scale=MatchLowercase}
  \defaultfontfeatures[\rmfamily]{Ligatures=TeX,Scale=1}
\fi
\usepackage{lmodern}
\ifPDFTeX\else
\fi
\IfFileExists{upquote.sty}{\usepackage{upquote}}{}
\IfFileExists{microtype.sty}{
  \usepackage[]{microtype}
  \UseMicrotypeSet[protrusion]{basicmath} 
}{}
\makeatletter
\@ifundefined{KOMAClassName}{
  \IfFileExists{parskip.sty}{%
    \usepackage{parskip}
  }{
    \setlength{\parindent}{0pt}
    \setlength{\parskip}{6pt plus 2pt minus 1pt}}
}{
  \KOMAoptions{parskip=half}}
\makeatother
\usepackage{xcolor}
\usepackage[margin=1in]{geometry}
\usepackage{color}
\usepackage{fancyvrb}

\DefineVerbatimEnvironment{Highlighting}{Verbatim}{commandchars=\\\{\}}
\usepackage{framed}
\definecolor{shadecolor}{RGB}{248,248,248}
\newenvironment{Shaded}{\begin{snugshade}}{\end{snugshade}}

\newcommand{\AttributeTok}[1]{\textcolor[rgb]{0.13,0.29,0.53}{#1}}

\newcommand{\ConstantTok}[1]{\textcolor[rgb]{0.56,0.35,0.01}{#1}}
\newcommand{\ControlFlowTok}[1]{\textcolor[rgb]{0.13,0.29,0.53}{\textbf{#1}}}

\newcommand{\DecValTok}[1]{\textcolor[rgb]{0.00,0.00,0.81}{#1}}

\newcommand{\FloatTok}[1]{\textcolor[rgb]{0.00,0.00,0.81}{#1}}
\newcommand{\FunctionTok}[1]{\textcolor[rgb]{0.13,0.29,0.53}{\textbf{#1}}}

\newcommand{\NormalTok}[1]{#1}

\newcommand{\OtherTok}[1]{\textcolor[rgb]{0.56,0.35,0.01}{#1}}

\newcommand{\SpecialCharTok}[1]{\textcolor[rgb]{0.81,0.36,0.00}{\textbf{#1}}}

\newcommand{\StringTok}[1]{\textcolor[rgb]{0.31,0.60,0.02}{#1}}

\usepackage{longtable,booktabs,array}
\usepackage{calc} 
\usepackage{etoolbox}
\makeatletter
\patchcmd\longtable{\par}{\if@noskipsec\mbox{}\fi\par}{}{}
\makeatother
\IfFileExists{footnotehyper.sty}{\usepackage{footnotehyper}}{\usepackage{footnote}}
\makesavenoteenv{longtable}
\usepackage{graphicx}
\makeatletter
\newsavebox\pandoc@box
\newcommand*\pandocbounded[1]{
  \sbox\pandoc@box{#1}%
  \Gscale@div\@tempa{\textheight}{\dimexpr\ht\pandoc@box+\dp\pandoc@box\relax}%
  \Gscale@div\@tempb{\linewidth}{\wd\pandoc@box}%
  \ifdim\@tempb\p@<\@tempa\p@\let\@tempa\@tempb\fi
  \ifdim\@tempa\p@<\p@\scalebox{\@tempa}{\usebox\pandoc@box}%
  \else\usebox{\pandoc@box}%
  \fi%
}
\def\fps@figure{htbp}
\makeatother
\providecommand{\tightlist}{%
  \setlength{\itemsep}{0pt}\setlength{\parskip}{0pt}}
\NewDocumentCommand\citeproctext{}{}

\makeatletter
 \let\@cite@ofmt\@firstofone
 \def\@biblabel#1{}
 \def\@cite#1#2{{#1\if@tempswa , #2\fi}}
\makeatother
\newlength{\cslhangindent}
\newlength{\csllabelwidth}
\newenvironment{CSLReferences}[2] 
 {\begin{list}{}{%
  \setlength{\itemindent}{0pt}
  \setlength{\leftmargin}{0pt}
  \setlength{\parsep}{0pt}
  \ifodd #1
   \setlength{\leftmargin}{\cslhangindent}
   \setlength{\itemindent}{-1\cslhangindent}
  \fi
  \setlength{\itemsep}{#2\baselineskip}}}
 {\end{list}}
\usepackage{calc}

\usepackage{tcolorbox}
\usepackage{amssymb}
\usepackage{yfonts}
\usepackage{bm}
\usepackage{titlesec}

\newtcolorbox{greybox}{
  colback=white,
  colframe=blue,
  coltext=black,
  boxsep=5pt,
  arc=4pt}

\usepackage{bookmark}
\IfFileExists{xurl.sty}{\usepackage{xurl}}{} 
\hypersetup{
  pdftitle={Convergence of SMACOF},
  pdfauthor={Jan de Leeuw - University of California Los Angeles},
  hidelinks,
  pdfcreator={LaTeX via pandoc}}

\title{Convergence of SMACOF}
\author{Jan de Leeuw - University of California Los Angeles}
\date{First created on August 18, 2019. Last update on July 17, 2024}

\begin{document}
\maketitle
\begin{abstract}
To study convergence of SMACOF we introduce a modification mSMACOF that rotates the configurations from each of the SMACOF iterations to principal components. This modification, called mSMACOF, has the same stress values as SMACOF in each iteration, but unlike SMACOF it produces a sequence of configurations that properly converges to a solution. We show that the modified algorithm can be implemented by iterating ordinary SMACOF to convergence, and then rotating the SMACOF solution to principal components. The speed of linear convergence of SMACOF and mSMACOF is the same, and is equal to the largest eigenvalue of the derivative of the Guttman transform, ignoring the trivial unit eigenvalues that result from rotational indeterminacy.
\end{abstract}

{
\setcounter{tocdepth}{3}
\tableofcontents
}
\textbf{Note:} \textbf{Note:} This is a working manuscript which may be updated.
All suggestions for improvement are welcome. All Rmd, tex,
html, pdf, R, and C files are in the public domain. Attribution will be
appreciated, but is not required. The files can be found at
\url{https://github.com/deleeuw/convergence}.

\section{Introduction}\label{introduction}

In (Euclidean, metric, least squares) multidimensional scaling we minimize the real valued loss function
\begin{equation}
\sigma(X)=\frac12\mathop{\sum}_{1\leq i<j\leq n} w_{ij}(\delta_{ij}-d_{ij}(X))^2,
\label{eq:stress}
\end{equation}
defined on \(\mathbb{R}^{n\times p}\), the space of all \(n\times p\) matrices. We follow Kruskal (1964) and call \(\sigma(X)\) the \emph{stress} of \emph{configuration} \(X\). Minimizing stress over p-dimensional configurations is the \emph{pMDS problem}.

In \eqref{eq:stress} the matrices of \emph{weights} \(W=\{w_{ij}\}\) and \emph{dissimilarities} \(\Delta=\{\delta_{ij}\}\) are symmetric, non-negative, and hollow (zero diagonal). The matrix-valued function \(D(X)=\{d_{ij}(X)\}\) contains Euclidean distances between the rows of the \emph{configuration} \(X\), which are the coordinates of \(n\) \emph{points} in \(\mathbb{R}^p\). Thus \(D(X)\) is also symmetric, non-negative, and hollow.

\section{Notation}\label{notation}

First some convenient notation, first introduced in De Leeuw (1977). Vector \(e_i\) has \(n\) elements, with element \(i\) equal to \(+1\), and all other elements zero. \(A_{ij}\) is the matrix \((e_i-e_j)(e_i-e_j)'\), which means elements \((i,i)\) and \((j,j)\) are equal to \(+1\), while \((i,j)\) and \((j,i)\) are \(-1\). Thus
\begin{equation}
d_{ij}^2(X)=(e_i-e_i)'XX'(e_i-e_j)=\text{tr}\ X'A_{ij}X=\text{tr}\ A_{ij}C,
\label{eq:anot}
\end{equation}
with \(C=XX'\).

We also define
\begin{equation}
V=\mathop{\sum}_{1\leq i<j\leq n}w_{ij}A_{ij},
\label{eq:vdef}
\end{equation}
and the matrix-valued function \(B\) with
\begin{equation}
B(X)=\mathop{\sum}_{d_{ij}(X)>0}w_{ij}\frac{\delta_{ij}}{d_{ij}(X)}A_{ij},
\label{eq:bdef}
\end{equation}
and \(B(X)=0\) if \(X=0\). If we assume, without loss of generality, that
\begin{equation}
\frac12\mathop{\sum}_{1\leq i<j\leq n}w_{ij}\delta_{ij}^2=1,
\label{eq:dnorm}
\end{equation}
then
\begin{equation}
\sigma(X)=1-\text{tr}\ X'B(X)X+\frac12\text{tr}\ X'VX.
\label{eq:S}
\end{equation}

We also suppose, without loss of generality, that \(W\) is \emph{irreducible}, so that the pMDS problem does not separate into a number of smaller pMDS problems. For symmetric matrices irreducibity means that we cannot find a permutation matrix \(\Pi\) such that \(\Pi'W\Pi\) is the direct sum of a number of smaller matrices.

\(V\) is symmetric with non-positive off-diagonal elements. It is doubly-centered (rows and columns add up to zero) and thus weakly diagonally dominant. It follows that it is positive semi-definite (see Varga (1962), section 1.5). Because of irreducibility it has rank \(n-1\), and the vectors in its null space are all proportional to \(e\), the vector with all elements equal to \(+1\). The matrix \(B(X)\) is also symmetric, positive semi-definite, and doubly-centered for each \(X\). It may not be irreducible, because for example \(B(0)=0\).

\section{SMACOF}\label{smacof}

Define the \emph{Guttman Transform} of configuration \(X\) as
\begin{equation}
\Gamma(X)=V^+B(X)X
\label{eq:guttman}
\end{equation}
The \emph{SMACOF algorithm} is
\begin{equation}
X^{(k+1)}=\Gamma(X^{(k)})=\Gamma^k(X^{0})
\label{eq:smacof}
\end{equation}
De Leeuw (1977) shows that the SMACOF iterations \eqref{eq:smacof} tend (in a specifc sense described later in this paper) to a fixed point of the Guttman transform, i.e.~to an \(X\) with \(\Gamma(X)=X\). If stress is differentiable at \(X\) then the fixed points of the Guttman transform are stationary points, where the derivative of stress vanishes. Also \(V^+B(X)X=X\) shows that the columns of \(X\) are eigenvectors of \(V^+B(X)\), with eigenvalues equal to one.

The algorithm was proposed by Guttman (1968), by differentiating stress and setting the derivatives equal to zero. This ignores the problem that stress is not differentiable if one or more distances are zero, and it also does not say anything about convergence of the algorithm. In De Leeuw (1977) a new derivation of the algorithm was given, using a majorization argument based on the Cauchy-Schwarz inequality. This make it possible to avoid problems with differentiability, and it leads to a simple convergence proof.

\section{Global Convergence of SMACOF}\label{global-convergence-of-smacof}

Following De Leeuw (1977) we also define
\begin{align}
\rho(X)&=\mathop{\sum}_{1\leq i<j\leq n} w_{ij}\delta_{ij}d_{ij}(X)=\text{tr}\ X'B(X)X,\\
\eta^2(X)&=\mathop{\sum}_{1\leq i<j\leq n} w_{ij}d_{ij}^2(X)=\text{tr}\ X'VX,
\end{align}
and
\begin{equation}
\lambda(X)=\frac{\rho(X)}{\eta(X)}.
\end{equation}
The main result in De Leeuw (1977) is:

\begin{enumerate}
\def\labelenumi{\arabic{enumi}.}
\tightlist
\item
  The SMACOF iterates \(X^{(k)}\) are in the compact set \(\eta(X)\leq 1\),
\item
  \(\{\rho(X^{k})\}, \{\eta^(X^{k})\}\) and \(\{\lambda(X^{k})\}\) are increasing sequences, with limits \(\rho_\infty,\eta_\infty\) and \(\lambda_\infty\), where \(\eta_\infty=\lambda_\infty=\sqrt{\rho_\infty}\).
\item
  \(\{\sigma(X^{k})\}\) is a decreasing sequence converging to \(\sigma_\infty=1-\eta^2_\infty\).
\item
  At any accumulation point \(X_\infty\) of \(\{\sigma(X^{k})\}\) we have \(\sigma(X_\infty)=\sigma_\infty\) and \(X_\infty=\Gamma(X_\infty)\).
\item
  \(\eta(X^{(k+1)}-X^{(k)})\rightarrow 0\) and thus either \(\{\sigma(X^{k})\}\) converges or the set of accumulation points of \(\{\sigma(X^{k})\}\) is a continuum.
\end{enumerate}

So, in words, the sequence of stress values converges monotonically. The sequence of configurations is asymptotically regular and has one or more accumulation points. Each accumulation point is a fixed point of the Guttman transform. All accumulation points have the same function value, which is also the limit of the stress sequence.

But it is important to emphasize that De Leeuw (1977) did not show that the configuration sequence \(\{X^{(k)}\}\) actually a Cauchy sequence, and converges to a configuration \(X_\infty\). This is basically because of the rotational invariance of the pMDS problem. If \(K\) is a rotation matrix, i.e.~\(K'K=KK'=I\), then \(d_{ij}(XK)=d_{ij}(X)\) for all \(i\) and \(j\), and thus \(\sigma(XK)=\sigma(X)\). If \(X\) is a fixed point of the Guttman tranform, then so is \(XK\).
Consequently there are no isolated fixed points, each fixed point is part of a nonlinear continuum of fixed points.

It is also of interest that De Leeuw (1984) showed that if \(X\) is a local minimizer of stress then \(d_{ij}(X)>0\) for all \(i\) and \(j\) such that \(w_{ij}\delta_{ij}>0\). Thus, if weights and dissimilarities are positive, stress is differentiable at a local minimum.
De Leeuw (1993) shows that stress has only a single local maximum at \(X=0\), and consequently the only possible stationary points are local minima and saddle points. It is shown by
De Leeuw (2019) that all fixed points of the Guttman transform which are not of full column rank are saddle points. Thus if \(X\) is a local minimizer of pMDS, then adding \(q\) zero columns to \(X\) produces a saddle point \([X\mid 0]\) for (p+q)MDS. At saddle points there are always directions in which stress can be decreased, and consequently the SMACOF algorithm with enough iterations will eventually get close to a continuum of local minima.

\section{Local Convergence of SMACOF}\label{local-convergence-of-smacof}

De Leeuw (1988) studies the local convergence of SMACOF, i.e.~the \emph{rate of convergence} of the SMACOF iterations. There is, however, a problem not adequately addressed in that article, which has quite a bit of hand-waiving, and some incorrect statements. If there is no convergence to a single configuration then the usual rate of convergence is not defined either.

We know the sequence \(\{\eta(X^{(k+1)}-X^{(k)})\}\) tends to zero. We can measure its rate of convergence by the \emph{ratio factor}
\begin{equation}
q_k=\frac{\eta(X^{(k+1)}-X^{(k)})}{\eta(X^{(k)}-X^{(k-1)})},
\end{equation}
and the \emph{root factor}
\begin{equation}
r_k=\eta(X^{(k+1)}-X^{(k)})^{1/k}.
\end{equation}
There is no guarantee that either \(\{q_k\}\) or \(\{r_k\}\) converges, but
\begin{equation}
\liminf_{k\rightarrow\infty} q_k\leq\liminf_{k\rightarrow\infty} r_k\leq\limsup_{k\rightarrow\infty} r_k\leq\limsup_{k\rightarrow\infty} q_k
\end{equation}
Thus if \(\lim_{k\rightarrow\infty} q_k\) exists then \(\lim_{k\rightarrow\infty} r_k\) exist and the two limit values are equal. See Rudin (1976), p.~68, theorem 3.37.

The rate of convergence of the SMACOF iterations at a fixed point \(X\) is computed in De Leeuw (1988) as the spectral norm \(\kappa(X)=\|\mathcal{D}\Gamma(X)\|_\infty\), i.e.~as the modulus of the largest eigenvalue of the derivative of the Guttman transform.
There are good reasons to compute this quantity. From Ostrovski's Theorem (Ostrowski (1973), theorem 22.1, p.~151 , Ortega and Rheinboldt (1970), theorem 10.1.3, p.~300) we know that if \(\kappa(X)<1\), then \(X\) is a \emph{point of attraction} of the SMACOF iteration. If we start close enough then the iterations will converge to \(X\). We also know (Ostrowski (1973), theorem 22.2, p.~152) that if \(\kappa(X)>1\) then \(X\) is a \emph{point of repulsion}, which means the iterations will diverge when started in a certain solid angle with \(X\) at its apex.

If we define, with Ortega and Rheinboldt (1970), chapter 9, for a sequence \(\{X^{(k)}\}\) converging to \(X\), the ratio and root convergence factors
\begin{equation}
Q_1(\{X^{(k)}\})=\limsup_{k\rightarrow\infty}\frac{\eta(X^{(k+1)}-X)}{\eta(X^{(k)}-X)},
\end{equation}
and
\begin{equation}
R_1(\{X^{(k)}\})=\limsup_{k\rightarrow\infty}\ \eta(X^{(k)}-X)^{1/k},
\end{equation}
then \(\kappa(X)<1\) implies \(R_1(X^{(k)})=\kappa(X)\) (Ortega and Rheinboldt (1970), theorem 10.1.4, p.~301). Alo \(R_1(X^{(k)})\) is independent of the norm we have chosen, which is \(\eta\) in the SMACOF case (Ortega and Rheinboldt (1970), theorem 9.9.2, p.~288). Because \(Q_1(X^{(k)})\) depends on the norm we can only assert that
\(Q_1(X^{(k)})\geq R_1(X^{(k)})=\kappa(X)\), although it is guaranteed in the SMACOF case that some norm exists for which there is equality (Ortega and Rheinboldt (1970), Notes and Remarks NR 10.1-5, p.~306).

There is a problem, however, with applying the Ostrowski and Ortega-Rheinboldt results in our case. Because of the invariance of distances under rotation we have \(\kappa(X)=1\), no matter what \(X\) is. And, basically for the same reason, we have not shown that the SMACOF iterations converge to a fixed point at all. De Leeuw (1988) on page 171 acknowledges the problem. In fact, he proposes a way around the problem in the proof of theorem 3 on page 175, but the proof is not very convincing, although convincing enough to get past the reviewers. I will try to be more specific and precise.

Observe that
\begin{equation}
\mathcal{D}\Gamma(X)=V^+\mathcal{D}^2\rho(X),
\label{eq:dgamma}
\end{equation}
and
\begin{equation}
\mathcal{D}^2\sigma(X)=V-\mathcal{D}^2\rho(X)=V(I-\mathcal{D}\Gamma(X)).
\label{eq:d2sigma}
\end{equation}
Because \(\rho\) is convex \(\mathcal{D}^2\rho(X)\) is symmetric and positive semi-definite.
The eigenvalues of \(\mathcal{D}\Gamma(X)\) are the generalized eigenvalues of the positive semi-definite matrix pair \((\mathcal{D}^2\rho(X),V)\).

An explicit formula for the derivatives of the Guttman transform, or equivalently for the second derivatives of stress, was first given by De Leeuw (1988). We write \(\mathcal{D}\Gamma(X)[Y]\) for the derivative evaluated at \(Y\). Partial derivatives can be obtained by choosing \(Y=e_ie_s'\). Of course we assume here that \(d_{ij}(X)>0\) for all
\(i<j\) with \(w_{ij}\delta_{ij}>0\).
\begin{equation}
\mathcal{D}\Gamma(X)[Y]=V^+\{B(X)Y-H(X,Y)X\},
\label{eq:dgammamat}
\end{equation}
with
\begin{equation}
H(X,Y)=\mathop{\sum}_{1\leq i<j\leq n}w_{ij}\frac{\delta_{ij}}{d_{ij}(X)}\frac{\text{tr}\ X'A_{ij}Y}{d_{ij}^2(X)}A_{ij}.
\label{eq:hdef}
\end{equation}
These formulas, together with equations \eqref{eq:dgamma} and \eqref{eq:d2sigma}, prove the main results in De Leeuw (1988).

\begin{enumerate}
\def\labelenumi{\arabic{enumi}.}
\tightlist
\item
  All eigenvalues of \(\mathcal{D}\Gamma(X)\) are non-negative.
\item
  \textbf{(Homogeneity)} \(X\) is an eigenvector of \(\mathcal{D}\Gamma(X)\) with eigenvalue zero.
\item
  \textbf{(Translation)} \(\mathcal{D}\Gamma(X)\) has at least \(p\) additional eigenvalues equal to zero, corresponding with eigenvectors of the form \(ee_s'\), where \(e\) has all elements equal to one and \(e_s\) has one of its elements equal to one and the others equal to zero.
\item
  \textbf{(Rotation)} If \(X\) is a fixed point of the Guttman transform then \(\mathcal{D}\Gamma(X)\) has at least \(\frac12 p(p-1)\) eigenvalues equal to one, corresponding with eigenvectors of the form \(XA\), with \(A\) anti-symmetric.
\item
  At a local minimum point \(X\) of stress all eigenvalues of \(\mathcal{D}\Gamma(X)\) are less than or equal to one. At a strict local minimum they are strictly less than one.
\item
  At a saddle point \(X\) of stress the largest eigenvalue of \(\mathcal{D}\Gamma(X)\) is larger than one.
\end{enumerate}

\section{Modified SMACOF}\label{modified-smacof}

To avoid the problems caused by a continuum of fixed points De Leeuw (1988) proposes to rotate each SMACOF iterate to a unique position, for example to principal components.
For any \(X\) with singular value decomposition \(X=K\Lambda L'\) we define \(\Pi(X)=K\Lambda=XL.\) Note that if one or more singular values are equal then
\(\Pi\) is not unique and becomes a point-to-set map. We can rotate within each of the spaces corresponding to multiple singular values, and thus we have not completely eliminated indeterminacy due to rotation. In the sequel we simply assume that the \(p\) singular values of \(X\) are different.

Now consider the iterations
\begin{equation}
\tilde X^{(k+1)}=\Pi(\Gamma(\tilde X^{(k)}))={(\Pi\Gamma)}^k(\tilde X^{0}).
\end{equation}
Let's call this modified SMACOF, or mSMACOF for short.

It looks initialy as if mSMACOF require a great deal of extra computation, but this is actually not the case. If \(M\) is a rotation matrix we have \(\Pi(XM)=\Pi(X)\) and \(\Gamma(XM)=\Gamma(X)M\). This implies that \(\Pi\Gamma\Pi(X)=\Pi\Gamma(X)\) for all \(X\). As a result the sequences \(\{X^{(k)}\}\) and \(\{\tilde X^{(k)}\}\) have a simple relationship. If we start the \(\{\tilde X^{(k)}\}\) sequence with \(X^{0}\) then for each \(k\) we have \(\sigma(\tilde X^{(k)})=\sigma(X^{(k)})\) and \(\tilde X^{(k)}=\Pi(X^{(k)})\). Thus the
two sequences of stress values are exactly the same for SMACOF and mSMACOF and the two configurations, and consequnetly also the accumulation points of the two sequences of configurations, just differ by a rotation. To get \(\tilde X^{(k)}\) we can just compute the SMACOF iterate \(X^{(k)}\) and rotate it to principal components.

The derivative of \(\Pi\Gamma\) is, by the chain rule,
\begin{equation}
\mathcal{D}\Pi\Gamma(X)=\mathcal{D}\Pi(\Gamma(X))\mathcal{D}\Gamma(X).
\end{equation}
After some computation we find
\begin{equation}
\mathcal{D}\Pi(X)[Y]=YL+K\Lambda M,
\end{equation}
where \(X=K\Lambda L'\) and \(M\) is the anti-symmetric matrix with off-diagonal elements
\begin{equation}
m_{ij}=-\frac{\lambda_iu_{ij}+\lambda_ju_{ji}}{\lambda_i^2-\lambda_j^2}
\end{equation}
with \(U=K'YL\).

If \(X\) satisfies \(\Pi(X)=X\) then \(L=I\) and thus \(\mathcal{D}\Pi(X)[Y]=Y+XM\) and \(U=K'Y\).
We proceed to compute the eigenvectors and eigenvalues of the Jacobian at an \(X\) with
\(\Pi(X)=X\).

\begin{enumerate}
\def\labelenumi{\arabic{enumi}.}
\tightlist
\item
  If \(Y=XA\) with \(A\) antisymmetric then \(M=-L'AL\) and \(\mathcal{D}\Pi(X)[Y]=0\).
  This corresponds with \(\frac12 p(p-1)\) zero eigenvalues
\item
  If \(Y=K\Lambda^{-1}A\) with \(A\) antisymmetric then \(M=0\). This gives \(\frac12 p(p-1)\)
  eigenvectors with eigenvalue one.
\item
  If \(Y=K\Lambda^{-1}D\) wth \(D\) diagonal then \(M=0\). This gives another \(p\) eigenvectors with eigenvalue one.
\item
  If \(Y=K_\perp S\) with \(K_\perp\) a basis for the orthogonal complement of \(X\) then \(M=0\) and we have another \(p(n-p)\) eigenvalues equal to one.
\end{enumerate}

This is a complete set of eigenvectors. It turns out all eigenvalues are equal to one, except for \(\frac12 p(p-1)\) zero eigenvalues. It follows that fixed points of
\(\Pi\Gamma\) are of the form \(\tilde X=XL\), with \(X\) a fixed point of \(\Gamma\) and \(L\) the rotation of \(X\) to principal components). Thus \(\Gamma(\tilde X)=\Gamma(X)L=XL=\tilde X\), and \(\tilde X\) is a fixed point of both \(\Gamma\) and \(\Pi\Gamma\). The eigenvalues of
\(\mathcal{D}\Pi\Gamma(\tilde X)\) are the same as the eigenvalues of \(\mathcal{D}\Gamma(X)\), except for the \(\frac12 p(p-1)\) eigenvalues equal to one, corresponding with \(XA\) for antisymmetric A, which are replaced by zero eigenvalues. The Ostrowski and Ortega-Rheinboldt
results now apply directly to \(\Pi\Gamma\). If there are no zero distances, if the singular values of \(X\) are different, and if the largest eigenvalue of \(\mathcal{D}\Pi\Gamma(X)\) is less than one, then \(X\) is a point of attraction, the SMACOF iterations converge to \(X\) if started sufficiently close to it (in the \emph{attraction ball} of \(X\)), and \(R_1(X)\) is equal to the largest eigenvalue.

\section{Software}\label{software}

The appendix has an ad-hoc version of the \texttt{smacof} function. Its arguments are

\begin{Shaded}
\begin{Highlighting}[]
\FunctionTok{args}\NormalTok{(smacof)}
\end{Highlighting}
\end{Shaded}

\begin{verbatim}
## function (delta, w, p = 2, xold = torgerson(delta, p), pca = FALSE, 
##     verbose = FALSE, eps = 1e-15, itmax = 10000) 
## NULL
\end{verbatim}

The function expects both dissimilarities and weights to be of class \texttt{dist}. The default initial configuration uses classical MDS (Torgerson (1958)). We iterate until either the distance \(\eta(X^{(k)}-X^{(k-1)})\) between successive configurations is less than \texttt{eps}, which has the ridiculously small default value of \ensuremath{10^{-15}}, or until the maximum number of iterations \texttt{itmax} is reached, which defaults to the ridiculously large default value of 10000. If \texttt{pca} is TRUE all iterates are rotated to principal components. If \texttt{verbose} is TRUE we print, for each iteration, the iteration number \(k\), the stress \(\sigma(X^{(k)})\), the change \(\eta(X^{(k)}-X^{(k-1)})\), and the root and ratio convergence factors.

The second file \texttt{derivative.R} computes the Jacobian of the iteration functions at the limit (i.e.~at the point where SMACOF stops). There are actually six functions in the file. The three functions \texttt{dGammaA,\ dPiA} and \texttt{dPiGammaA} use the analytical expressions we have derived earlier for \(\mathcal{D}\Gamma(X), \mathcal{D}\Pi(X)\) and \(\mathcal{D}\Pi\Gamma(X).\) The corresponding functions \texttt{dGammaN,\ dPiN} and \texttt{dPiGammaN} numerically compute the Jacobian using the \texttt{jacobian} function from the \texttt{numDeriv} package (Gilbert and Varadhan (2019)). They are basically used to check the formulas.

\section{Examples}\label{examples}

\subsection{Ekman}\label{ekman}

Time for a numerical example. We will use the color similarity data of Ekman (1954), taken from the SMACOF package (De Leeuw and Mair (2009)).
We transform Ekman's similarities \(s_{ij}\) to dissimilarities \(\delta_{ij}\) using \(\delta_{ij}=(1-s_{ij})^3\), and we use unit weights in \(W\).

SMACOF requires 51 iterations for convergence. The minimum of stress is 0.0110248119,
the root factor is equal to 0.5074583707 and the ratio factor is 0.5478850001. The eigenvalues of \(\mathcal{D}\Gamma(X)\) are

\begin{verbatim}
## 0.999999999999999 
## 0.538510668196407 
## 0.532498554224166 
## 0.529669334191043 
## 0.525541097754551 
## 0.519602718218807 
## 0.516533687841054 
## 0.513727908691608 
## 0.510295310409166 
## 0.506357695934493 
## 0.495649713446001 
## 0.481518780073304 
## 0.469548625536451 
## 0.445038589020916 
## 0.410479252654484 
## 0.394284315478172 
## 0.357670750114585 
## 0.348395413045201 
## 0.330341477528788 
## 0.313520384427863 
## 0.287598015841956 
## 0.280399104121623 
## 0.267278474596759 
## 0.247631495462991 
## 0.216576009083047 
## 0.000000000000000 
## 0.000000000000000 
## 0.000000000000000
\end{verbatim}

As an aside, the Ekman example (with this particular transformation of the similarities) is special, because the two-dimensional solution is actually the global minimum over all configurations (i.e.~the global mnimum of pMDS for all \(1\leq p\leq n\)). As shown in De Leeuw (2014) this follows from the fact that the two unit eigenvalues of \(V^+B(X)\) are actually its two largest eigenvalues, and consequently \(X\) is the solution of the convex full-dimensional nMDS relaxation of the pMDS problem (De Leeuw, Groenen, and Mair (2016)). The eigenvalues of \(V^+B(X)\) are

\begin{verbatim}
## 1.000000000000000 
## 0.999999999999999 
## 0.923497086367286 
## 0.907901212970888 
## 0.862936584878895 
## 0.852692003103344 
## 0.829803620857356 
## 0.814556167662381 
## 0.793238576338502 
## 0.791651722456648 
## 0.786442678118769 
## 0.747679475705024 
## 0.728268247434340 
## 0.000000000000000
\end{verbatim}

We now set \texttt{pca=TRUE} and run mSMACOF. It requires 51 iterations for convergence. The minimum of stress is 0.0110248119, and the root factor is equal to 0.5075280657 and the ratio factor is 0.549074338. The eigenvalues
of \(\mathcal{D}\Pi\Gamma(X)\) at the fixed point \(X\) are

\begin{verbatim}
## 0.538510668196406 
## 0.532498554224167 
## 0.529669334191043 
## 0.525541097754552 
## 0.519602718218807 
## 0.516533687841054 
## 0.513727908691610 
## 0.510295310409167 
## 0.506357695934492 
## 0.495649713446000 
## 0.481518780073304 
## 0.469548625536451 
## 0.445038589020916 
## 0.410479252654485 
## 0.394284315478172 
## 0.357670750114583 
## 0.348395413045200 
## 0.330341477528788 
## 0.313520384427863 
## 0.287598015841956 
## 0.280399104121623 
## 0.267278474596759 
## 0.247631495462991 
## 0.216576009083047 
## 0.000000000000000 
## 0.000000000000000 
## 0.000000000000000 
## 0.000000000000000
\end{verbatim}

They are equal to the eigenvalues of \(\mathcal{D}\Gamma(X)\) and the root convergence factor
\(R_1(\{X^{(k)}\})\) is 0.5385106682.

\subsection{De Gruijter}\label{de-gruijter}

The second example are dissimilarties between nine Dutch political parties, collected a long time ago by De Gruijter (1967).

\begin{verbatim}
##       KVP PvdA  VVD  ARP  CHU  CPN  PSP   BP
## PvdA 2.63                                   
## VVD  2.27 3.72                              
## ARP  1.60 2.64 2.46                         
## CHU  1.80 3.22 1.97 0.20                    
## CPN  4.54 2.12 5.13 4.84 4.80               
## PSP  3.73 1.59 4.55 3.73 4.08 1.08          
## BP   4.18 4.22 3.90 4.28 3.96 3.34 3.88     
## D66  3.17 2.47 1.67 3.13 3.04 4.42 3.36 4.36
\end{verbatim}

We analyze the data with SMACOF using three dimensions.

SMACOF requires 778 iterations for convergence. The minimum of stress is 0.003442194, the root factor is equal to 0.9565703351, and the ratio factor is 0.9584004108. The eigenvalues of \(\mathcal{D}\Gamma(X)\) are

\begin{verbatim}
## 1.000000000000000 
## 1.000000000000000 
## 1.000000000000000 
## 0.965505429805660 
## 0.940592046981168 
## 0.919047686446446 
## 0.863993920924432 
## 0.822263696226349 
## 0.810020971414307 
## 0.771947328045071 
## 0.728539213663602 
## 0.712621684571534 
## 0.659318624263221 
## 0.638485365688917 
## 0.624552464148481 
## 0.616085432872672 
## 0.557480497708779 
## 0.501954186928525 
## 0.458630667850014 
## 0.450598367846592 
## 0.355243400669833 
## 0.309186479174062 
## 0.247708397109091 
## 0.000000000000002 
## 0.000000000000002 
## 0.000000000000001 
## 0.000000000000000
\end{verbatim}

In this case there is no guarantee that we have the three-dimensional global minimum. The unit eigenvalues of the matrix \(V^+B(X)\) are not the three largest ones.

\begin{verbatim}
## 1.079524009371954 
## 1.032606649163672 
## 1.000000000000000 
## 1.000000000000000 
## 1.000000000000000 
## 0.986706272372899 
## 0.971839080877692 
## 0.906211919383163 
## 0.000000000000001
\end{verbatim}

mSMACOF in three dimensions requires 779 iterations for convergence. The minimum of stress is 0.003442194,
the root factor is equal to 0.9566140798, and the ratio factor is 0.89001211. The eigenvalues of \(\mathcal{D}\Pi\Gamma(X)\) are

\begin{verbatim}
## 0.965505429805661 
## 0.940592046981168 
## 0.919047686446444 
## 0.863993920924433 
## 0.822263696226350 
## 0.810020971414308 
## 0.771947328045072 
## 0.728539213663603 
## 0.712621684571535 
## 0.659318624263222 
## 0.638485365688918 
## 0.624552464148482 
## 0.616085432872671 
## 0.557480497708778 
## 0.501954186928525 
## 0.458630667850015 
## 0.450598367846592 
## 0.355243400669832 
## 0.309186479174061 
## 0.247708397109091 
## 0.000000000000001 
## 0.000000000000001 
## 0.000000000000000 
## 0.000000000000000 
## 0.000000000000000 
## 0.000000000000000 
## 0.000000000000000
\end{verbatim}

Again, as expected, the eigenvalues for SMACOF and mSMACOF are the same and the root convergence factor is the spectral norm of the mSMACOF derivative, which is 0.9655054298.

\section{Conclusion}\label{conclusion}

From a practical point of view there is no need to ever use modified SMACOF. We only use the Guttman transform, and compute the largest eigenvalue of its derivative at the solution \(X\), ignoring the \(\frac12 p(p-1)\) trivial unit eigenvalues. That largest eigenvalue, if it is strictly smaller than one, is the root convergence factor. In that case SMACOF can be said to converge to \(\Pi(X)\), which is the SMACOF solution rotated to principal components.

\section{Appendix: Code}\label{appendix-code}

\subsection{smacof.R}\label{smacof.r}

\begin{Shaded}
\begin{Highlighting}[]
\NormalTok{torgerson }\OtherTok{\textless{}{-}} \ControlFlowTok{function}\NormalTok{ (delta, }\AttributeTok{p =} \DecValTok{2}\NormalTok{) \{}
\NormalTok{  h }\OtherTok{\textless{}{-}} \FunctionTok{as.matrix}\NormalTok{(delta }\SpecialCharTok{\^{}} \DecValTok{2}\NormalTok{)}
\NormalTok{  n }\OtherTok{\textless{}{-}} \FunctionTok{nrow}\NormalTok{ (h)}
\NormalTok{  j }\OtherTok{\textless{}{-}} \FunctionTok{diag}\NormalTok{ (n) }\SpecialCharTok{{-}}\NormalTok{ (}\DecValTok{1} \SpecialCharTok{/}\NormalTok{ n)}
\NormalTok{  h }\OtherTok{\textless{}{-}} \SpecialCharTok{{-}}\NormalTok{(j }\SpecialCharTok{\%*\%}\NormalTok{ h }\SpecialCharTok{\%*\%}\NormalTok{ j) }\SpecialCharTok{/} \DecValTok{2}
\NormalTok{  e }\OtherTok{\textless{}{-}} \FunctionTok{eigen}\NormalTok{ (h)}
  \FunctionTok{return}\NormalTok{ (e}\SpecialCharTok{$}\NormalTok{vectors[, }\DecValTok{1}\SpecialCharTok{:}\NormalTok{p] }\SpecialCharTok{\%*\%} \FunctionTok{diag}\NormalTok{(}\FunctionTok{sqrt}\NormalTok{ (e}\SpecialCharTok{$}\NormalTok{values[}\DecValTok{1}\SpecialCharTok{:}\NormalTok{p])))}
\NormalTok{\}}

\NormalTok{smacof }\OtherTok{\textless{}{-}}
  \ControlFlowTok{function}\NormalTok{ (delta,}
\NormalTok{            w,}
            \AttributeTok{p =} \DecValTok{2}\NormalTok{,}
            \AttributeTok{xold =} \FunctionTok{torgerson}\NormalTok{ (delta, p),}
            \AttributeTok{pca =} \ConstantTok{FALSE}\NormalTok{,}
            \AttributeTok{verbose =} \ConstantTok{FALSE}\NormalTok{,}
            \AttributeTok{eps =} \FloatTok{1e{-}15}\NormalTok{,}
            \AttributeTok{itmax =} \DecValTok{10000}\NormalTok{) \{}
\NormalTok{    n }\OtherTok{\textless{}{-}} \FunctionTok{round}\NormalTok{ ((}\DecValTok{1} \SpecialCharTok{+} \FunctionTok{sqrt}\NormalTok{ (}\DecValTok{1} \SpecialCharTok{+} \DecValTok{8} \SpecialCharTok{*} \FunctionTok{length}\NormalTok{ (delta))) }\SpecialCharTok{/} \DecValTok{2}\NormalTok{)}
\NormalTok{    v }\OtherTok{\textless{}{-}} \SpecialCharTok{{-}}\FunctionTok{as.matrix}\NormalTok{(w)}
    \FunctionTok{diag}\NormalTok{(v) }\OtherTok{\textless{}{-}} \SpecialCharTok{{-}}\FunctionTok{rowSums}\NormalTok{(v)}
\NormalTok{    vinv }\OtherTok{\textless{}{-}} \FunctionTok{solve}\NormalTok{(v }\SpecialCharTok{+}\NormalTok{ (}\DecValTok{1} \SpecialCharTok{/}\NormalTok{ n)) }\SpecialCharTok{{-}}\NormalTok{ (}\DecValTok{1} \SpecialCharTok{/}\NormalTok{ n)}
\NormalTok{    delta }\OtherTok{\textless{}{-}}\NormalTok{ delta }\SpecialCharTok{/} \FunctionTok{sqrt}\NormalTok{ (}\FunctionTok{sum}\NormalTok{ (w }\SpecialCharTok{*}\NormalTok{ delta }\SpecialCharTok{\^{}} \DecValTok{2}\NormalTok{) }\SpecialCharTok{/} \DecValTok{2}\NormalTok{)}
\NormalTok{    dold }\OtherTok{\textless{}{-}} \FunctionTok{dist}\NormalTok{ (xold)}
\NormalTok{    sold }\OtherTok{\textless{}{-}} \FunctionTok{sum}\NormalTok{ (w }\SpecialCharTok{*}\NormalTok{ (delta }\SpecialCharTok{{-}}\NormalTok{ dold) }\SpecialCharTok{\^{}} \DecValTok{2}\NormalTok{) }\SpecialCharTok{/} \DecValTok{2}
\NormalTok{    eold }\OtherTok{\textless{}{-}} \ConstantTok{Inf}
\NormalTok{    itel }\OtherTok{\textless{}{-}} \DecValTok{1}
    \ControlFlowTok{repeat}\NormalTok{ \{}
\NormalTok{      b }\OtherTok{\textless{}{-}} \FunctionTok{as.matrix}\NormalTok{ (}\SpecialCharTok{{-}}\NormalTok{w }\SpecialCharTok{*}\NormalTok{ delta }\SpecialCharTok{/}\NormalTok{ dold)}
      \FunctionTok{diag}\NormalTok{ (b) }\OtherTok{\textless{}{-}} \SpecialCharTok{{-}}\FunctionTok{rowSums}\NormalTok{(b)}
\NormalTok{      xnew }\OtherTok{\textless{}{-}}\NormalTok{ vinv }\SpecialCharTok{\%*\%}\NormalTok{ b }\SpecialCharTok{\%*\%}\NormalTok{ xold}
      \ControlFlowTok{if}\NormalTok{ (pca) \{}
\NormalTok{        xsvd }\OtherTok{\textless{}{-}} \FunctionTok{svd}\NormalTok{ (xnew)}
\NormalTok{        xnew }\OtherTok{\textless{}{-}}\NormalTok{ xnew }\SpecialCharTok{\%*\%}\NormalTok{ xsvd}\SpecialCharTok{$}\NormalTok{v}
\NormalTok{      \}}
\NormalTok{      dnew }\OtherTok{\textless{}{-}} \FunctionTok{dist}\NormalTok{ (xnew)}
\NormalTok{      snew }\OtherTok{\textless{}{-}} \FunctionTok{sum}\NormalTok{ (w }\SpecialCharTok{*}\NormalTok{ (delta }\SpecialCharTok{{-}}\NormalTok{ dnew) }\SpecialCharTok{\^{}} \DecValTok{2}\NormalTok{) }\SpecialCharTok{/} \DecValTok{2}
\NormalTok{      enew }\OtherTok{\textless{}{-}} \FunctionTok{sqrt}\NormalTok{ (}\FunctionTok{sum}\NormalTok{ (v }\SpecialCharTok{*} \FunctionTok{tcrossprod}\NormalTok{(xold }\SpecialCharTok{{-}}\NormalTok{ xnew)))}
\NormalTok{      rnew }\OtherTok{\textless{}{-}}\NormalTok{ enew }\SpecialCharTok{\^{}}\NormalTok{ (}\DecValTok{1} \SpecialCharTok{/}\NormalTok{ itel)}
\NormalTok{      qnew }\OtherTok{\textless{}{-}}\NormalTok{ enew }\SpecialCharTok{/}\NormalTok{ eold}
      \ControlFlowTok{if}\NormalTok{ (verbose) \{}
        \FunctionTok{cat}\NormalTok{(}
          \StringTok{"itel "}\NormalTok{,}
          \FunctionTok{formatC}\NormalTok{(itel, }\AttributeTok{digits =} \DecValTok{4}\NormalTok{, }\AttributeTok{format =} \StringTok{"d"}\NormalTok{),}
          \StringTok{"loss "}\NormalTok{,}
          \FunctionTok{formatC}\NormalTok{(snew, }\AttributeTok{digits =} \DecValTok{15}\NormalTok{, }\AttributeTok{format =} \StringTok{"f"}\NormalTok{),}
          \StringTok{"chan "}\NormalTok{,}
          \FunctionTok{formatC}\NormalTok{(enew, }\AttributeTok{digits =} \DecValTok{15}\NormalTok{, }\AttributeTok{format =} \StringTok{"f"}\NormalTok{),}
          \StringTok{"rcnf "}\NormalTok{,}
          \FunctionTok{formatC}\NormalTok{(rnew, }\AttributeTok{digits =} \DecValTok{15}\NormalTok{, }\AttributeTok{format =} \StringTok{"f"}\NormalTok{),}
          \StringTok{"qcnf "}\NormalTok{,}
          \FunctionTok{formatC}\NormalTok{(qnew, }\AttributeTok{digits =} \DecValTok{15}\NormalTok{, }\AttributeTok{format =} \StringTok{"f"}\NormalTok{),}
          \StringTok{"}\SpecialCharTok{\textbackslash{}n}\StringTok{"}
\NormalTok{        )}
\NormalTok{      \}}
      \ControlFlowTok{if}\NormalTok{ ((enew }\SpecialCharTok{\textless{}}\NormalTok{ eps) }\SpecialCharTok{||}\NormalTok{ (itel }\SpecialCharTok{==}\NormalTok{ itmax))}
        \ControlFlowTok{break}
\NormalTok{      xold }\OtherTok{\textless{}{-}}\NormalTok{ xnew}
\NormalTok{      dold }\OtherTok{\textless{}{-}}\NormalTok{ dnew}
\NormalTok{      sold }\OtherTok{\textless{}{-}}\NormalTok{ snew}
\NormalTok{      eold }\OtherTok{\textless{}{-}}\NormalTok{ enew}
\NormalTok{      itel }\OtherTok{\textless{}{-}}\NormalTok{ itel }\SpecialCharTok{+} \DecValTok{1}
\NormalTok{    \}}
\NormalTok{    out }\OtherTok{\textless{}{-}}
      \FunctionTok{list}\NormalTok{ (}
        \AttributeTok{itel =}\NormalTok{ itel,}
        \AttributeTok{x =}\NormalTok{ xnew,}
        \AttributeTok{s =}\NormalTok{ snew,}
        \AttributeTok{q =}\NormalTok{ qnew,}
        \AttributeTok{r =}\NormalTok{ rnew,}
        \AttributeTok{b =}\NormalTok{ vinv }\SpecialCharTok{\%*\%}\NormalTok{ b}
\NormalTok{      )}
    \FunctionTok{return}\NormalTok{ (out)}
\NormalTok{  \}}
\end{Highlighting}
\end{Shaded}

\subsection{derivative.R}\label{derivative.r}

\begin{Shaded}
\begin{Highlighting}[]
\NormalTok{dGammaA }\OtherTok{\textless{}{-}} \ControlFlowTok{function}\NormalTok{ (x, delta, w) \{}
\NormalTok{  n }\OtherTok{\textless{}{-}} \FunctionTok{nrow}\NormalTok{ (x)}
\NormalTok{  p }\OtherTok{\textless{}{-}} \FunctionTok{ncol}\NormalTok{ (x)}
\NormalTok{  h }\OtherTok{\textless{}{-}} \FunctionTok{matrix}\NormalTok{ (}\DecValTok{0}\NormalTok{, n }\SpecialCharTok{*}\NormalTok{ p, n }\SpecialCharTok{*}\NormalTok{ p)}
\NormalTok{  d }\OtherTok{\textless{}{-}} \FunctionTok{dist}\NormalTok{ (x)}
\NormalTok{  delta }\OtherTok{\textless{}{-}}\NormalTok{ delta }\SpecialCharTok{/} \FunctionTok{sqrt}\NormalTok{ (}\FunctionTok{sum}\NormalTok{ (w }\SpecialCharTok{*}\NormalTok{ delta }\SpecialCharTok{\^{}} \DecValTok{2}\NormalTok{) }\SpecialCharTok{/} \DecValTok{2}\NormalTok{)}
\NormalTok{  dmat }\OtherTok{\textless{}{-}} \FunctionTok{as.matrix}\NormalTok{ (d)}
\NormalTok{  wmat }\OtherTok{\textless{}{-}} \FunctionTok{as.matrix}\NormalTok{ (w)}
\NormalTok{  emat }\OtherTok{\textless{}{-}} \FunctionTok{as.matrix}\NormalTok{ (delta)}
\NormalTok{  v }\OtherTok{\textless{}{-}} \SpecialCharTok{{-}}\NormalTok{wmat}
  \FunctionTok{diag}\NormalTok{(v) }\OtherTok{\textless{}{-}} \SpecialCharTok{{-}}\FunctionTok{rowSums}\NormalTok{ (v)}
\NormalTok{  b }\OtherTok{\textless{}{-}} \FunctionTok{as.matrix}\NormalTok{(}\SpecialCharTok{{-}}\NormalTok{w }\SpecialCharTok{*}\NormalTok{ delta }\SpecialCharTok{/}\NormalTok{ d)}
  \FunctionTok{diag}\NormalTok{(b) }\OtherTok{\textless{}{-}} \SpecialCharTok{{-}}\FunctionTok{rowSums}\NormalTok{ (b)}
\NormalTok{  vinv }\OtherTok{\textless{}{-}} \FunctionTok{solve}\NormalTok{(v }\SpecialCharTok{+}\NormalTok{ (}\DecValTok{1} \SpecialCharTok{/}\NormalTok{ n)) }\SpecialCharTok{{-}}\NormalTok{ (}\DecValTok{1} \SpecialCharTok{/}\NormalTok{ n)}
  \ControlFlowTok{for}\NormalTok{ (s }\ControlFlowTok{in} \DecValTok{1}\SpecialCharTok{:}\NormalTok{p) \{}
    \ControlFlowTok{for}\NormalTok{ (t }\ControlFlowTok{in} \DecValTok{1}\SpecialCharTok{:}\NormalTok{p) \{}
\NormalTok{      gst }\OtherTok{\textless{}{-}} \FunctionTok{matrix}\NormalTok{ (}\DecValTok{0}\NormalTok{, n, n)}
      \ControlFlowTok{for}\NormalTok{ (i }\ControlFlowTok{in} \DecValTok{1}\SpecialCharTok{:}\NormalTok{n) \{}
        \ControlFlowTok{for}\NormalTok{ (j }\ControlFlowTok{in} \DecValTok{1}\SpecialCharTok{:}\NormalTok{n) \{}
          \ControlFlowTok{if}\NormalTok{ (i }\SpecialCharTok{==}\NormalTok{ j)}
            \ControlFlowTok{next}
\NormalTok{          gs }\OtherTok{\textless{}{-}}\NormalTok{ x[i, s] }\SpecialCharTok{{-}}\NormalTok{ x[j, s]}
\NormalTok{          gt }\OtherTok{\textless{}{-}}\NormalTok{ x[i, t] }\SpecialCharTok{{-}}\NormalTok{ x[j, t]}
\NormalTok{          gst[i, j] }\OtherTok{\textless{}{-}}
            \SpecialCharTok{{-}}\NormalTok{wmat[i, j] }\SpecialCharTok{*}\NormalTok{ emat[i, j] }\SpecialCharTok{*}\NormalTok{ gs }\SpecialCharTok{*}\NormalTok{ gt }\SpecialCharTok{/}\NormalTok{ (dmat[i, j] }\SpecialCharTok{\^{}} \DecValTok{3}\NormalTok{)}
\NormalTok{        \}}
\NormalTok{      \}}
      \FunctionTok{diag}\NormalTok{(gst) }\OtherTok{\textless{}{-}} \SpecialCharTok{{-}}\FunctionTok{rowSums}\NormalTok{ (gst)}
\NormalTok{      h[(s }\SpecialCharTok{{-}} \DecValTok{1}\NormalTok{) }\SpecialCharTok{*}\NormalTok{ n }\SpecialCharTok{+} \DecValTok{1}\SpecialCharTok{:}\NormalTok{n, (t }\SpecialCharTok{{-}} \DecValTok{1}\NormalTok{) }\SpecialCharTok{*}\NormalTok{ n }\SpecialCharTok{+} \DecValTok{1}\SpecialCharTok{:}\NormalTok{n] }\OtherTok{\textless{}{-}} \SpecialCharTok{{-}}\NormalTok{vinv }\SpecialCharTok{\%*\%}\NormalTok{ gst}
\NormalTok{    \}}
\NormalTok{  \}}
  \ControlFlowTok{for}\NormalTok{ (s }\ControlFlowTok{in} \DecValTok{1}\SpecialCharTok{:}\NormalTok{p) \{}
\NormalTok{    kn }\OtherTok{\textless{}{-}}\NormalTok{ (s }\SpecialCharTok{{-}} \DecValTok{1}\NormalTok{) }\SpecialCharTok{*}\NormalTok{ n }\SpecialCharTok{+} \DecValTok{1}\SpecialCharTok{:}\NormalTok{n}
\NormalTok{    h[kn, kn] }\OtherTok{\textless{}{-}}\NormalTok{ h[kn, kn] }\SpecialCharTok{+}\NormalTok{ vinv }\SpecialCharTok{\%*\%}\NormalTok{ b}
\NormalTok{  \}}
  \FunctionTok{return}\NormalTok{ (h)}
\NormalTok{\}}

\NormalTok{dPiA }\OtherTok{\textless{}{-}} \ControlFlowTok{function}\NormalTok{ (x) \{}
\NormalTok{  n }\OtherTok{\textless{}{-}} \FunctionTok{nrow}\NormalTok{ (x)}
\NormalTok{  p }\OtherTok{\textless{}{-}} \FunctionTok{ncol}\NormalTok{ (x)}
\NormalTok{  sx }\OtherTok{\textless{}{-}} \FunctionTok{svd}\NormalTok{ (x)}
\NormalTok{  xu }\OtherTok{\textless{}{-}}\NormalTok{ sx}\SpecialCharTok{$}\NormalTok{u}
\NormalTok{  xv }\OtherTok{\textless{}{-}}\NormalTok{ sx}\SpecialCharTok{$}\NormalTok{v}
\NormalTok{  xd }\OtherTok{\textless{}{-}}\NormalTok{ sx}\SpecialCharTok{$}\NormalTok{d}
\NormalTok{  h }\OtherTok{\textless{}{-}} \FunctionTok{matrix}\NormalTok{ (}\DecValTok{0}\NormalTok{, n }\SpecialCharTok{*}\NormalTok{ p, n }\SpecialCharTok{*}\NormalTok{ p)}
  \ControlFlowTok{for}\NormalTok{ (s }\ControlFlowTok{in} \DecValTok{1}\SpecialCharTok{:}\NormalTok{p) \{}
    \ControlFlowTok{for}\NormalTok{ (i }\ControlFlowTok{in} \DecValTok{1}\SpecialCharTok{:}\NormalTok{n) \{}
\NormalTok{      ir }\OtherTok{\textless{}{-}}\NormalTok{ (s }\SpecialCharTok{{-}} \DecValTok{1}\NormalTok{) }\SpecialCharTok{*}\NormalTok{ n }\SpecialCharTok{+}\NormalTok{ i}
\NormalTok{      e }\OtherTok{\textless{}{-}} \FunctionTok{matrix}\NormalTok{ (}\DecValTok{0}\NormalTok{, n, p)}
\NormalTok{      m }\OtherTok{\textless{}{-}} \FunctionTok{matrix}\NormalTok{ (}\DecValTok{0}\NormalTok{, p, p)}
\NormalTok{      e[i, s] }\OtherTok{\textless{}{-}} \DecValTok{1}
\NormalTok{      u }\OtherTok{\textless{}{-}} \FunctionTok{crossprod}\NormalTok{ (xu, e }\SpecialCharTok{\%*\%}\NormalTok{ xv)}
      \ControlFlowTok{for}\NormalTok{ (k }\ControlFlowTok{in} \DecValTok{1}\SpecialCharTok{:}\NormalTok{p) \{}
        \ControlFlowTok{for}\NormalTok{ (l }\ControlFlowTok{in} \DecValTok{1}\SpecialCharTok{:}\NormalTok{p) \{}
          \ControlFlowTok{if}\NormalTok{ (k }\SpecialCharTok{==}\NormalTok{ l)}
            \ControlFlowTok{next}
\NormalTok{          m[k, l] }\OtherTok{\textless{}{-}}
\NormalTok{            (xd[k] }\SpecialCharTok{*}\NormalTok{ u[k, l] }\SpecialCharTok{+}\NormalTok{ xd[l] }\SpecialCharTok{*}\NormalTok{ u[l, k]) }\SpecialCharTok{/}\NormalTok{ (xd[k] }\SpecialCharTok{\^{}} \DecValTok{2} \SpecialCharTok{{-}}\NormalTok{ xd[l] }\SpecialCharTok{\^{}} \DecValTok{2}\NormalTok{)}
\NormalTok{        \}}
\NormalTok{      \}}
\NormalTok{      h[, ir] }\OtherTok{\textless{}{-}} \FunctionTok{as.vector}\NormalTok{ (e }\SpecialCharTok{\%*\%}\NormalTok{ xv }\SpecialCharTok{{-}}\NormalTok{ xu }\SpecialCharTok{\%*\%} \FunctionTok{diag}\NormalTok{ (xd) }\SpecialCharTok{\%*\%}\NormalTok{ m)}
\NormalTok{    \}}
\NormalTok{  \}}
  \FunctionTok{return}\NormalTok{ (h)}
\NormalTok{\}}

\NormalTok{dGammaN }\OtherTok{\textless{}{-}} \ControlFlowTok{function}\NormalTok{ (x, delta, w) \{}
\NormalTok{  n }\OtherTok{\textless{}{-}} \FunctionTok{nrow}\NormalTok{ (x)}
\NormalTok{  p }\OtherTok{\textless{}{-}} \FunctionTok{ncol}\NormalTok{ (x)}
\NormalTok{  delta }\OtherTok{\textless{}{-}}\NormalTok{ delta }\SpecialCharTok{/} \FunctionTok{sqrt}\NormalTok{ (}\FunctionTok{sum}\NormalTok{ (w }\SpecialCharTok{*}\NormalTok{ delta }\SpecialCharTok{\^{}} \DecValTok{2}\NormalTok{) }\SpecialCharTok{/} \DecValTok{2}\NormalTok{)}
\NormalTok{  v }\OtherTok{\textless{}{-}} \SpecialCharTok{{-}} \FunctionTok{as.matrix}\NormalTok{ (w)}
  \FunctionTok{diag}\NormalTok{ (v) }\OtherTok{\textless{}{-}} \SpecialCharTok{{-}} \FunctionTok{rowSums}\NormalTok{(v)}
\NormalTok{  vinv }\OtherTok{\textless{}{-}} \FunctionTok{solve}\NormalTok{ (v }\SpecialCharTok{+}\NormalTok{ (}\DecValTok{1} \SpecialCharTok{/}\NormalTok{ n)) }\SpecialCharTok{{-}}\NormalTok{ (}\DecValTok{1} \SpecialCharTok{/}\NormalTok{ n)}
\NormalTok{  guttman }\OtherTok{\textless{}{-}} \ControlFlowTok{function}\NormalTok{ (x) \{}
\NormalTok{    d }\OtherTok{\textless{}{-}} \FunctionTok{dist}\NormalTok{ (}\FunctionTok{matrix}\NormalTok{ (x, n, p))}
\NormalTok{    b }\OtherTok{\textless{}{-}} \SpecialCharTok{{-}} \FunctionTok{as.matrix}\NormalTok{ (w }\SpecialCharTok{*}\NormalTok{ delta }\SpecialCharTok{/}\NormalTok{ d)}
    \FunctionTok{diag}\NormalTok{ (b) }\OtherTok{\textless{}{-}} \SpecialCharTok{{-}} \FunctionTok{rowSums}\NormalTok{ (b)}
\NormalTok{    z }\OtherTok{\textless{}{-}}\NormalTok{ vinv }\SpecialCharTok{\%*\%}\NormalTok{ b }\SpecialCharTok{\%*\%} \FunctionTok{matrix}\NormalTok{ (x, n, p)}
    \FunctionTok{return}\NormalTok{ (}\FunctionTok{as.vector}\NormalTok{ (z))}
\NormalTok{  \}}
  \FunctionTok{return}\NormalTok{ (}\FunctionTok{jacobian}\NormalTok{ (guttman, }\FunctionTok{as.vector}\NormalTok{(x)))}
\NormalTok{\}}

\NormalTok{dPiGammaN }\OtherTok{\textless{}{-}} \ControlFlowTok{function}\NormalTok{ (x, delta, w) \{}
\NormalTok{  n }\OtherTok{\textless{}{-}} \FunctionTok{nrow}\NormalTok{ (x)}
\NormalTok{  p }\OtherTok{\textless{}{-}} \FunctionTok{ncol}\NormalTok{ (x)}
\NormalTok{  delta }\OtherTok{\textless{}{-}}\NormalTok{ delta }\SpecialCharTok{/} \FunctionTok{sqrt}\NormalTok{ (}\FunctionTok{sum}\NormalTok{ (w }\SpecialCharTok{*}\NormalTok{ delta }\SpecialCharTok{\^{}} \DecValTok{2}\NormalTok{) }\SpecialCharTok{/} \DecValTok{2}\NormalTok{)}
\NormalTok{  v }\OtherTok{\textless{}{-}} \SpecialCharTok{{-}} \FunctionTok{as.matrix}\NormalTok{ (w)}
  \FunctionTok{diag}\NormalTok{ (v) }\OtherTok{\textless{}{-}} \SpecialCharTok{{-}} \FunctionTok{rowSums}\NormalTok{(v)}
\NormalTok{  vinv }\OtherTok{\textless{}{-}} \FunctionTok{solve}\NormalTok{ (v }\SpecialCharTok{+}\NormalTok{ (}\DecValTok{1} \SpecialCharTok{/}\NormalTok{ n)) }\SpecialCharTok{{-}}\NormalTok{ (}\DecValTok{1} \SpecialCharTok{/}\NormalTok{ n)}
\NormalTok{  guttman }\OtherTok{\textless{}{-}} \ControlFlowTok{function}\NormalTok{ (x) \{}
\NormalTok{    d }\OtherTok{\textless{}{-}} \FunctionTok{dist}\NormalTok{ (}\FunctionTok{matrix}\NormalTok{ (x, n, p))}
\NormalTok{    b }\OtherTok{\textless{}{-}} \SpecialCharTok{{-}} \FunctionTok{as.matrix}\NormalTok{ (w }\SpecialCharTok{*}\NormalTok{ delta }\SpecialCharTok{/}\NormalTok{ d)}
    \FunctionTok{diag}\NormalTok{ (b) }\OtherTok{\textless{}{-}} \SpecialCharTok{{-}} \FunctionTok{rowSums}\NormalTok{ (b)}
\NormalTok{    z }\OtherTok{\textless{}{-}}\NormalTok{ vinv }\SpecialCharTok{\%*\%}\NormalTok{ b }\SpecialCharTok{\%*\%} \FunctionTok{matrix}\NormalTok{ (x, n, p)}
\NormalTok{    z }\OtherTok{\textless{}{-}}\NormalTok{ z }\SpecialCharTok{\%*\%} \FunctionTok{svd}\NormalTok{(z)}\SpecialCharTok{$}\NormalTok{v}
    \FunctionTok{return}\NormalTok{ (}\FunctionTok{as.vector}\NormalTok{ (z))}
\NormalTok{  \}}
  \FunctionTok{return}\NormalTok{ (}\FunctionTok{jacobian}\NormalTok{ (guttman, }\FunctionTok{as.vector}\NormalTok{(x)))}
\NormalTok{\}}

\NormalTok{dPiGammaA }\OtherTok{\textless{}{-}} \ControlFlowTok{function}\NormalTok{ (x, delta, w) \{}
\NormalTok{  n }\OtherTok{\textless{}{-}} \FunctionTok{nrow}\NormalTok{ (x)}
\NormalTok{  guttman }\OtherTok{\textless{}{-}} \ControlFlowTok{function}\NormalTok{ (x, delta, w) \{}
\NormalTok{    d }\OtherTok{\textless{}{-}} \FunctionTok{dist}\NormalTok{ (x)}
\NormalTok{    b }\OtherTok{\textless{}{-}} \FunctionTok{as.matrix}\NormalTok{ (}\SpecialCharTok{{-}}\NormalTok{ w }\SpecialCharTok{*}\NormalTok{ delta }\SpecialCharTok{/}\NormalTok{ d)}
    \FunctionTok{diag}\NormalTok{ (b) }\OtherTok{\textless{}{-}} \SpecialCharTok{{-}} \FunctionTok{rowSums}\NormalTok{ (b)}
\NormalTok{    v }\OtherTok{\textless{}{-}} \SpecialCharTok{{-}} \FunctionTok{as.matrix}\NormalTok{ (w)}
    \FunctionTok{diag}\NormalTok{ (v) }\OtherTok{\textless{}{-}} \SpecialCharTok{{-}} \FunctionTok{rowSums}\NormalTok{ (v)}
\NormalTok{    vinv }\OtherTok{\textless{}{-}} \FunctionTok{solve}\NormalTok{(v }\SpecialCharTok{+}\NormalTok{ (}\DecValTok{1} \SpecialCharTok{/}\NormalTok{ n)) }\SpecialCharTok{{-}}\NormalTok{ (}\DecValTok{1} \SpecialCharTok{/}\NormalTok{ n)}
    \FunctionTok{return}\NormalTok{ (vinv }\SpecialCharTok{\%*\%}\NormalTok{ b }\SpecialCharTok{\%*\%}\NormalTok{ x)}
\NormalTok{  \}}
\NormalTok{  h }\OtherTok{\textless{}{-}} \FunctionTok{dGammaA}\NormalTok{ (x, delta, w)}
\NormalTok{  g }\OtherTok{\textless{}{-}} \FunctionTok{dPiA}\NormalTok{ (}\FunctionTok{guttman}\NormalTok{ (x, delta, w))}
  \FunctionTok{return}\NormalTok{ (g }\SpecialCharTok{\%*\%}\NormalTok{ h)}
\NormalTok{\}}
\end{Highlighting}
\end{Shaded}

\section*{References}\label{references}
\addcontentsline{toc}{section}{References}

\phantomsection\label{refs}

\end{document}